\documentclass{PoS}
\pdfoutput=1

\usepackage{mathtools}
\usepackage{graphicx, amssymb, amsfonts, amsmath}

\title{Approaching Conformality}

\ShortTitle{Approaching Conformality}

\author{Maria Paola Lombardo\\
        INFN-Laboratori Nazionali di Frascati\\
        E-mail: \email{mariapaola.lombardo@lnf.infn.it}}

\author{Kohtaroh Miura\\
        Kobayashi-Maskawa Institute for the Origin of Particles and the Universe(KMI), Nagoya University\\
        E-mail: \email{miura@yukawa.kyoto-u.ac.jp}}

\author{\speaker{Tiago Nunes da Silva}\\
        Van Swinderen Institute, University of Groningen\\
        E-mail: \email{t.j.nunes@rug.nl}}

\author{Elisabetta Pallante\\
        Van Swinderen Institute, University of Groningen\\
        E-mail: \email{e.pallante@rug.nl}}

\abstract{We investigate the preconformal region of the phase diagram of $SU(3)$ theories with fundamental flavors. We have simulated $SU(3)$ theories with six and eight fundamental flavors at volumes $32^3\times 64$. We use the generated configurations to measure the string tension $\sigma$ and the $w_0$ scale setting quantity extracted from the gradient flow. We show preliminary results on the ratios $T_c/\sqrt{\sigma}$ and $T_cw_0$. We compare them to the behavior obtained at smaller $N_f$ and discuss the implications of our results.}

\FullConference{The 32nd International Symposium on Lattice Field Theory,\\
		23-28 June, 2014\\
		Columbia University New York, NY}

\begin{document}

\section{Introduction}

Conformal invariance is expected to be restored in non-Abelian gauge theories with a large flavor content before asymptotic freedom is lost, due to the emergency of a nontrivial infrared fixed point (IRFP)\cite{Caswell:1974gg}. This emergence of conformality at large number of flavors $N_f$ plays an important role in the shape of the phase diagram of such theories. Relevant questions are to map the region of the parameter space where the conformal behavior happens and to determine which type of dynamics leads to it. The understanding of the phase diagram is important in the context of comprehending the fundamental structure of strong interactions, and it is also phenomenologically relevant as it can provide a framework for model builders working on Beyond the Standard Model scenarios. Because of that, it has received much attention, notably within the lattice community, given that the strongly coupled nature of the problem asks for a nonperturbative approach. Several groups are exploring the phase diagram for various gauge groups and different fermion representations. In this work we will focus on many-flavored QCD-like theories, i.e., $SU(N_c=3)$ theories with several species of flavors in the fundamental representation.

\begin{figure}[ht]
\centering
\includegraphics[width=.45\textwidth]{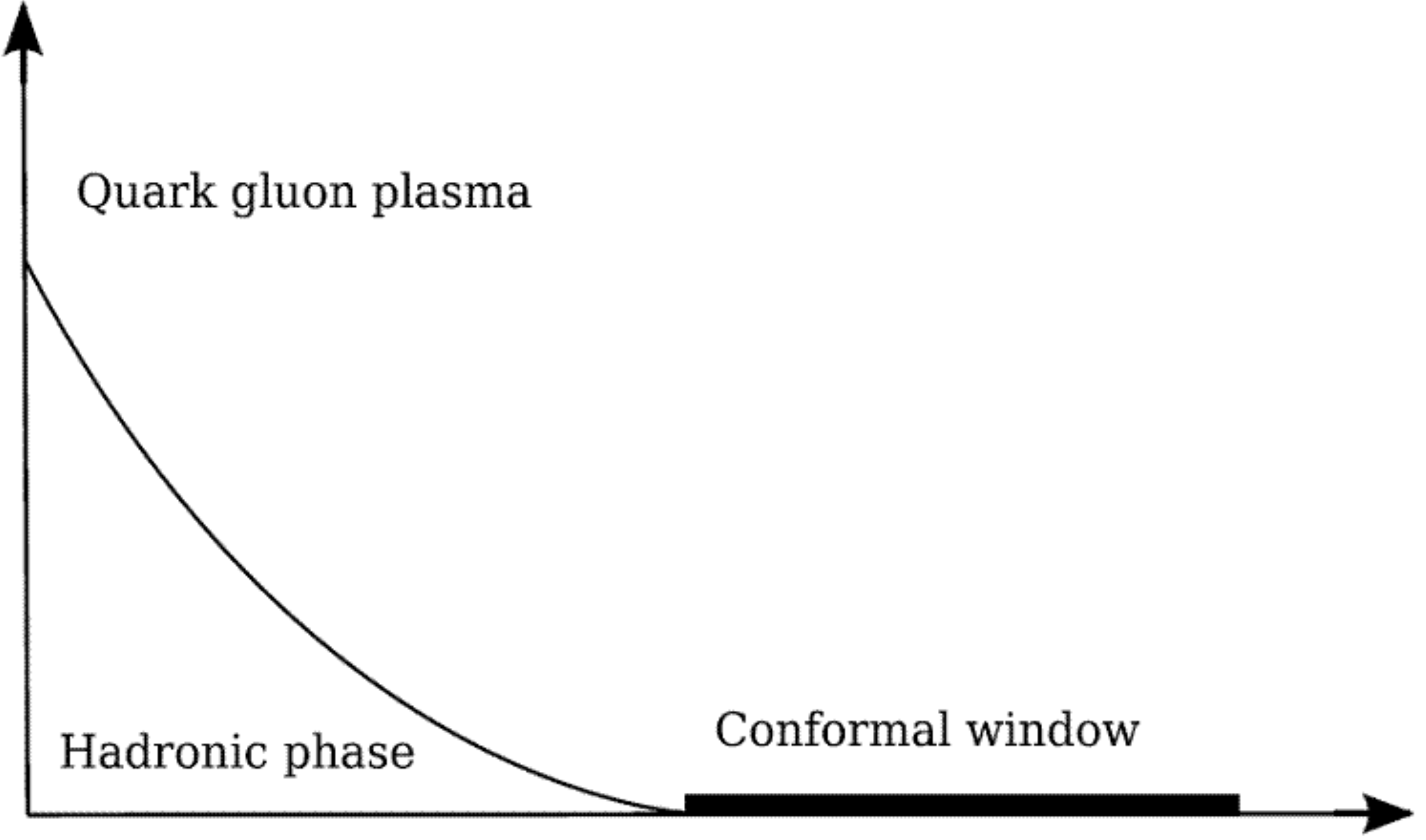}
\caption{ \label{fig:PhaseDia} Conjectured picture of the phase diagram in the $T\times N_f$ plane. In the region of small $N_f$ the chiral phase boundary separates the hadronic (chirally broken and confined) sector at low temperatures of QCD-like theories from the QGP (chirally symmetric and deconfined) sector at high temperatures. For $N_f$ large enough, but still before asymptotic freedom is lost, the conformal window opens and the theories exhibit restored chiral symmetry even at $T=0$. Theories living inside the window are conformal at the IFRP.}
\end{figure}

Activity on this particular family of theories has been very intensive. For a recent review of our own results, see \cite{Lombardo:2014mda}. The usual strategy followed is to try to establish if a given theory, individually, presents evidence of conformality. Such an approach, however, can be quite elusive for theories lying close to the critical number of flavors $N_f^c$ that separates theories that do realize conformality from theories that don't. In this work we follow an alternative approach. For small $N_f$ these theories present a finite temperature transition line separating a hadronic\,---\,chirally broken and confined\,---\,sector at low (and zero) temperature from the quark gluon plasma (QGP)\,---\,chirally restored and deconfined\,---\,sector at high temperatures. On the conjectured phase diagram depicted in Fig~\ref{fig:PhaseDia} this is represented by the chiral phase boundary line, which is expected to go to zero at $N_f^c$ in the conformal window scenario of conformal symmetry restoration \cite{Miransky:1996pd,Appelquist:1998rb}.  Inspired by the work done by Holger and Gies \cite{Braun:2010qs}, we aim at obtaining a direct evidence for $N_f^c$ and probing precursory effects of conformality by investigating the vanishing of the chiral phase boundary for values of $N_f$ close and smaller than $N_f^c$. 

The challenge lies in the fact that by changing the number of flavors we obtain fundamentally different theories, so that a comparison between the critical temperature obtained for each theory requires a proper normalization by a common scale. In order to properly disentangle the infrared behavior of the chiral phase boundary this scale must be ultraviolet. The behavior of $T_c$ measured on a reference UV scale is one example of a more general aspect of the pre-conformal dynamics itself: the theory should develop different dynamical scales, which should be reflected by the behavior of adimensional ratios when $N_f \to N_f^c$. In the remainder of this work we present our simulations setup, some preliminary results and a discussion and outlook. 

\section{Simulation Setup}

In this work, we study SU(3) theories with $N_f = 6$ and $N_f = 8$ degenerate fundamental flavors. We measure two normalization scales: the string tension $\sigma$ and the $w_0$ scale setting quantity proposed by the BMW collaboration in \cite{Borsanyi:2012zs}, and obtain the ratios $T_c/\sqrt{\sigma}$ and $T_cw_0$, with $T_c$, in lattice units, determined in \cite{Miura:2012zqa}. For each value of $N_f$, we performed zero temperature simulations at volume $32^3 \times 64$ and fixed bare mass $am = 0.02$ at the critical $\beta$s of the finite temperature simulations with $N_t = 6,8$ from \cite{Miura:2012zqa}.  This choice of $\beta$ values is summarized in Table \ref{table-betas}.    
\begin{table}[htb]
\centering
\begin{tabular}{| c | c | c |}
	\hline
	 & $N_t = 6$ & $N_t = 8$  \\ \hline
	 $N_f = 6$ & $\beta = 5.025 $ & $\beta = 5.200$ \\
	 $N_f = 8$ & $\beta =4.1125 $ & $\beta = 4.275$\\
	 \hline
\end{tabular}
\caption{\label{table-betas} Choice of bare couplings used in this work. Each $\beta$ value corresponds to the critical value of the bare coupling for $SU(3)$ with the indicated $N_f$ obtained from previous simulations with the corresponding $N_t$.}
\end{table}

We have utilized the MILC suite to simulate the four ensembles corresponding to the choices above with a one-loop Symanzik and tadpole improved gauge action and the Asqtad fermion action. Measurements of the chiral condensate and of the plaquette were conducted on the fly to monitor the thermalization of the system. The tadpole improvement factor $u_0$ was tuned to satisfy the relation $u_0 = \langle U_{C_p}\rangle^{1/4}$. After thermalization of the ensembles, we have saved lattice configurations separated by an interval of approximately 2 molecular dynamics time units (MDTU). On these configurations, we measure the Wilson Loops $W_{r,t}$, where $r$ and $t$ denote the spatial and temporal coordinates, respectively. From the Wilson Loops,  the heavy-quark potential is obtained using the relation $W_{r,t} = C(r)e^{-tV(r)}$.
The value of the string tension is finally obtained by fitting the lattice results for the potential with the ansatz
\begin{equation}
V(r) = V_0 - \frac{\alpha}{r}+\sigma r.
\label{eq:pot-ansatz}
\end{equation}

The configurations were also used for measuring the $w_0$ quantity from the gradient flow. The flow is defined by the equations
\begin{gather}
\dot{B}_\mu = D_\nu G_{\nu\mu} \\
G_{\mu\nu} = \partial_\mu B_\nu - \partial_\nu B_\mu + \left[ B_\mu, B_\nu \right], \hspace{0.5cm} D_\mu = \partial_\mu + \left[B_\mu,\cdot\right]
\end{gather}
where a dot represents differentiation over the flow time $t$ (which has units of inverse mass-squared). Calculating the flow on the lattice is equivalent to solving the flow equation:
\begin{equation}
\dot{V}_t=Z(V_t)V_t, \hspace{1cm} V_0=U.
\label{eq:flow}
\end{equation}
Here $V_t$ and $U$ represent the gauge links at flow time $t$ and the original gauge links, respectively. In \cite{Luscher:2010iy}, the author uses the Wilson action, so that $Z(V_t)$ is the derivative of the plaquette action and the flow is called the Wilson flow. The original prescription by Luscher \cite{Luscher:2009eq, Luscher:2010iy} was to compute the flow and measure the quantity $t^2 \langle E(t) \rangle$ as a function of $t$ until it reaches 0.3. The quantity $\langle E(t) \rangle$ is the expectation value of the continuum-like action density $G_{\mu \nu}^a (t) G_{\mu \nu}^a (t) / 4$. Here $G_{\mu \nu}^a (t)$ is a lattice version of the chromoelectric field-strength tensor at flow time $t$. The scale $t_0$ is given by the time $t$ at which $t^2 \langle E(t) \rangle = 0.3$. More recently, the BMW collaboration has proposed the use of a related scale: $w_0$ \cite{Borsanyi:2012zs}. Their idea is to use as a {\it proper observable} the quantity
\begin{equation}
W(t) \equiv t \frac{d}{dt} \left\{t^2 \langle E(t) \rangle \right\}
\end{equation}
and define the scale $w_0$ by the condition $W(t)|_{t=w_0^2} = 0.3$ This procedure carries all the benefits from the method proposed by Luscher and has as advantage over the scale $t_0$ the fact that it is less susceptible to discretization effects in the region of small $t \sim a^2$. In this work, we extract the quantity $w_0$ both from the Wilson and Symanzik flows. 

\section{Results}
\begin{figure}[ht]
\centering
\includegraphics[width=.45\textwidth]{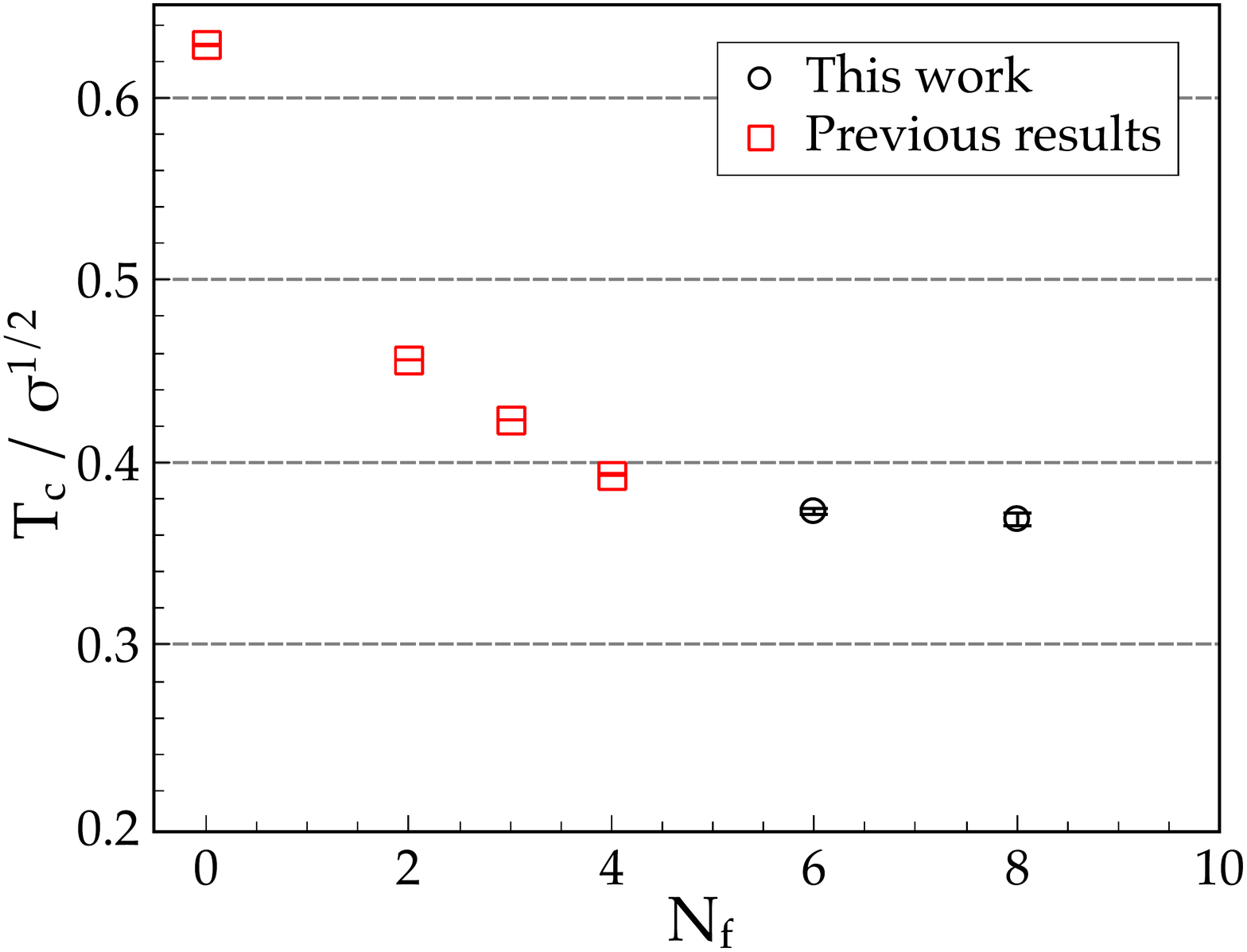}
\hfill
\includegraphics[width=.45\textwidth,origin=c]{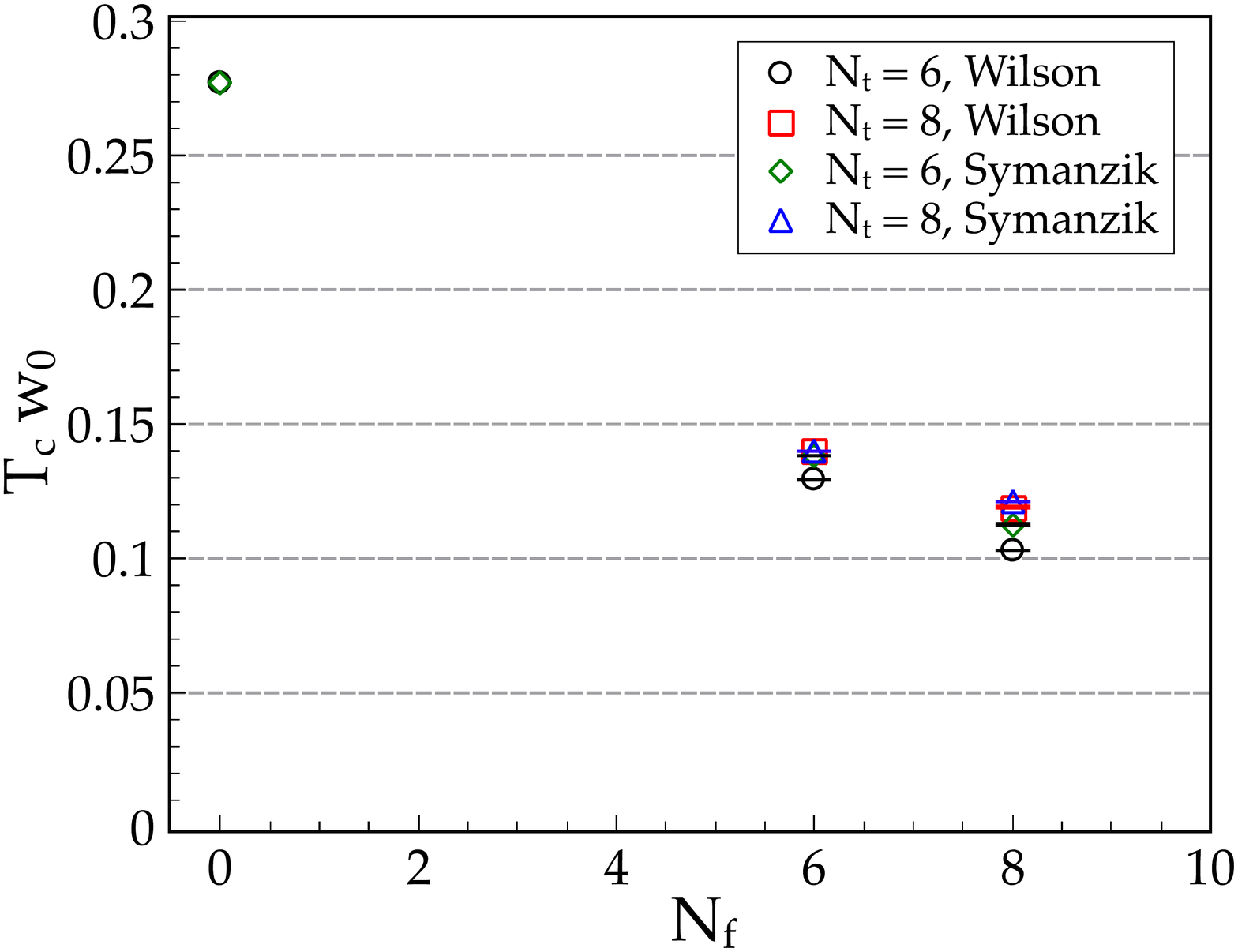}
\caption{ The ratios $T_c/\sqrt{\sigma}$ (left) and $T_cw_0$ (right). The plot on the left contains our new results (black circles) and results presented previously in the literature (red squares) \cite{Laermann:1996xn, Karsch:2000kv, Engels:1996ag}. }
\label{fig:tcRatios}
\end{figure}

We start by presenting our results for the string tension. This is an update to the results presented in \cite{Miura:2014aqa}. Here, we only present the results for our ensembles $N_f=6, \beta = 5.025$ and $N_f = 8, \beta = 4.275$. Figure~\ref{fig:potentials} show the heavy quark potential obtained from these ensembles, and the best fit obtained. The fits yield the following values for the string tension:
\begin{figure}[ht]
\centering
\includegraphics[width=.45\textwidth]{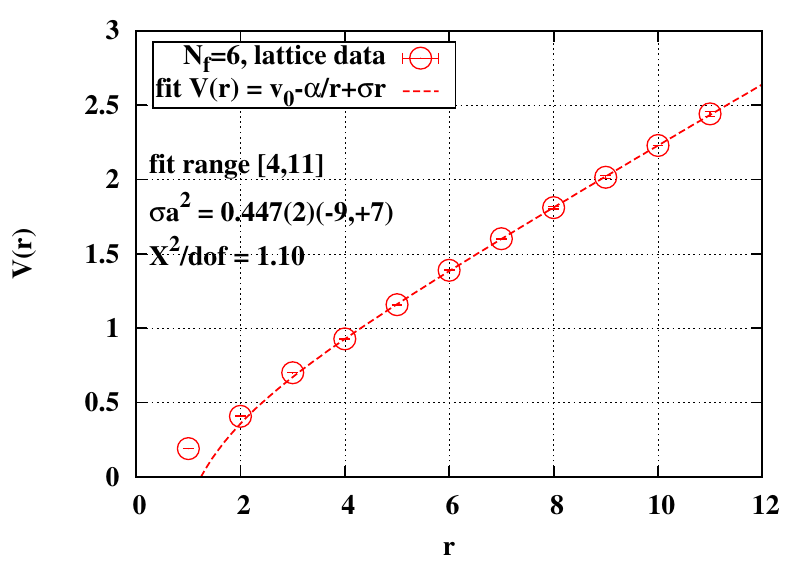}
\hfill
\includegraphics[width=.45\textwidth,origin=c]{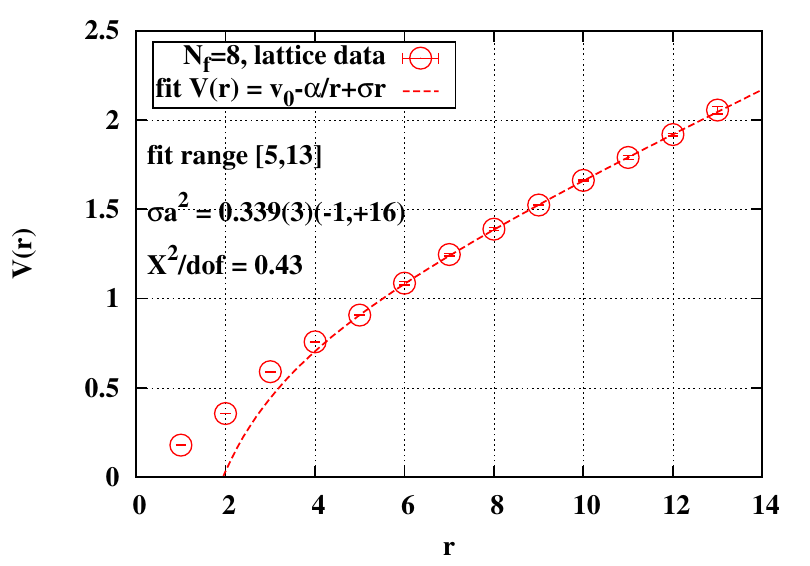}
\caption{The heavy-quark potential obtained for our ensembles $N_f=6, \beta = 5.025$ (left) and $N_f = 8, \beta = 4.275$(right). Results were obtained using APE and time-link smearing. Dashed lines indicate the best fit.}
\label{fig:potentials}
\end{figure}
\begin{equation}
\sigma a^2=\begin{cases} 0.447(2)(-9,+7), \hspace{1cm} N_f = 6, \beta = 5.025, \\ 0.339(3)(-1,+16), \hspace{0.8cm} N_f = 8, \beta = 4.275. \end{cases}
\end{equation}
The data points for the potential at small $r$  suffer from smearing artifacts. To address this issue, we have explored several fitting ranges for the potential data to the ansatz of Eq.~(\ref{eq:pot-ansatz}), and cite as central value and associated statistical error those coming from the largest fitting range that still yield a reasonable $\chi^2/\text{d.o.f.}$. The systematical error associated with the choice of fitting range is also quoted in the results. The corresponding results for $T_c/\sqrt{\sigma}$ are given by
\begin{equation}
T_c/\sqrt{\sigma} =\begin{cases} 0.373(2)(+5,-6), \hspace{1cm} N_f = 6, \beta = 5.025, \\ 0.369(4)(+1,-5), \hspace{1cm} N_f = 8, \beta = 4.275. \end{cases}
\end{equation}
The measurements obtained for the ratio $T_c/\sqrt{\sigma}$ are plotted in Figure~\ref{fig:tcRatios} (left) together with results for other values of $N_f$ which were already present in the literature \cite{Laermann:1996xn, Karsch:2000kv, Engels:1996ag}. The decreasing trend observed at small $N_f$ becomes milder with increasing $N_f$ and $T_c/\sqrt{\sigma}$ flattens, with no sign of vanishing before asymptotic freedom is lost. This is not surprising. The string tension probes the confining nature of the system and the long distance contributions to the potential, and therefore could be expected not to be an UV quantity and to vanish at the onset of the conformal window, where also $T_c$ vanishes. This would render it impossible to use it as a common reference scale to explore the chiral phase boundary. It is also possible that, because the finite quark mass used in the simulation breaks the conformality, both $T_c$ and $\sigma$ remain finite even for $N_f > N_f^c$. 

\begin{figure}[ht]
\centering
\includegraphics[width=.45\textwidth]{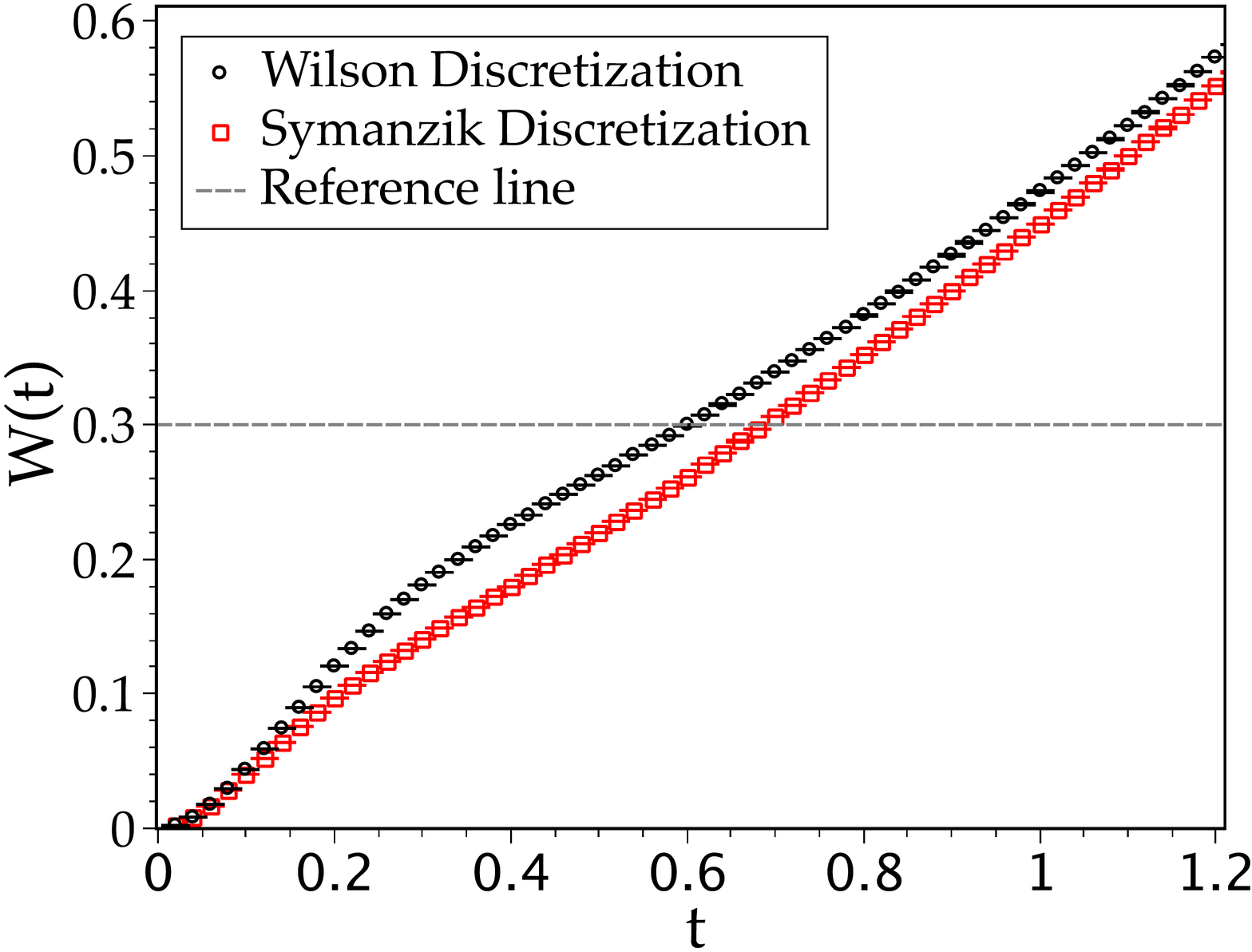}
\hfill
\includegraphics[width=.45\textwidth,origin=c]{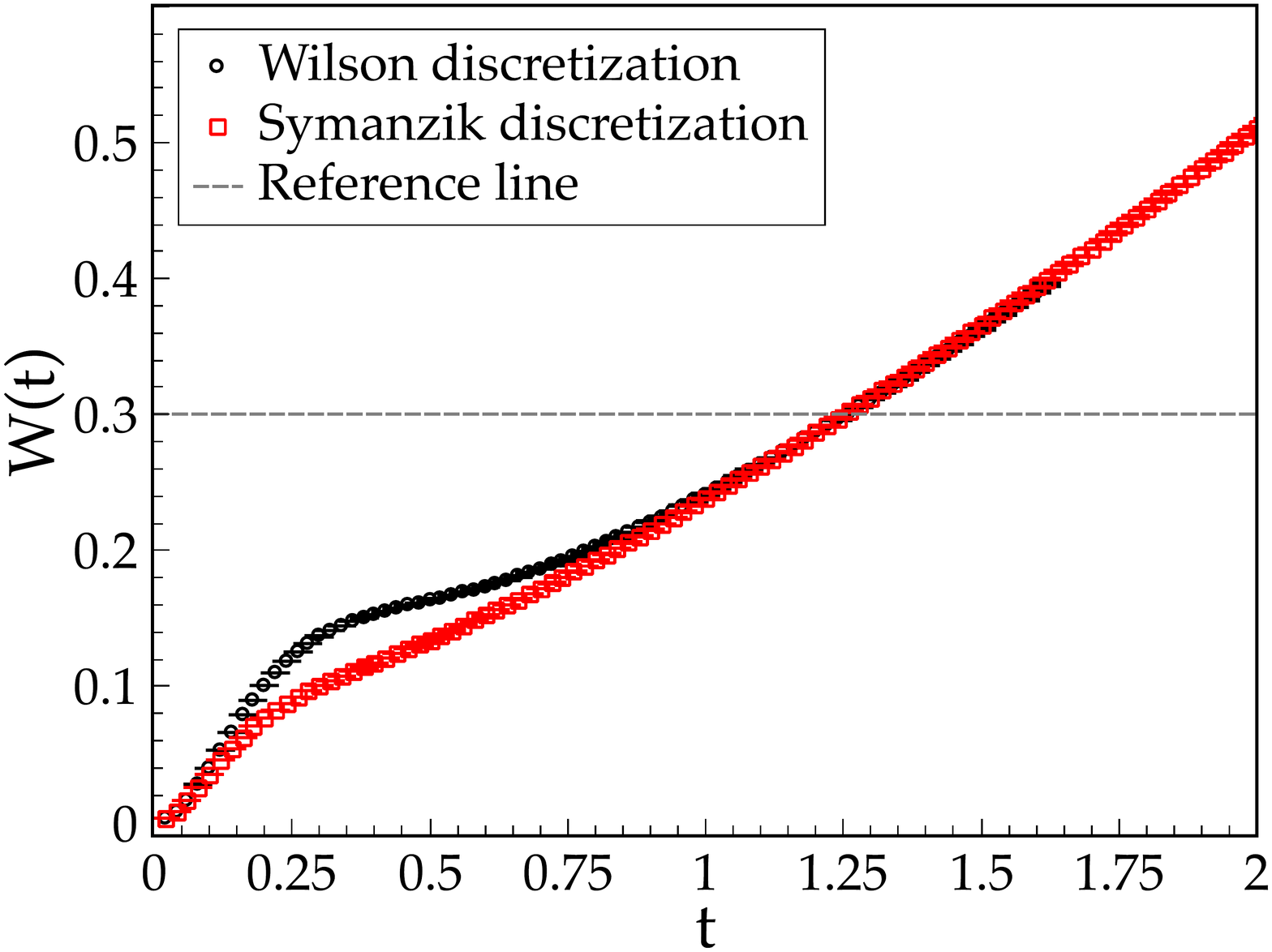}
\caption{ Plots of $W(t)$ for the ensembles $N_f = 6, \beta = 5.025$ (left) and $N_f=6, \beta = 5.200$ (right). The flows were measured with both Wilson (black circles) and Symanzik (red squares) discretizations. }
\label{fig:flows-nf6}
\end{figure}

\begin{figure}[ht]
\centering
\includegraphics[width=.45\textwidth]{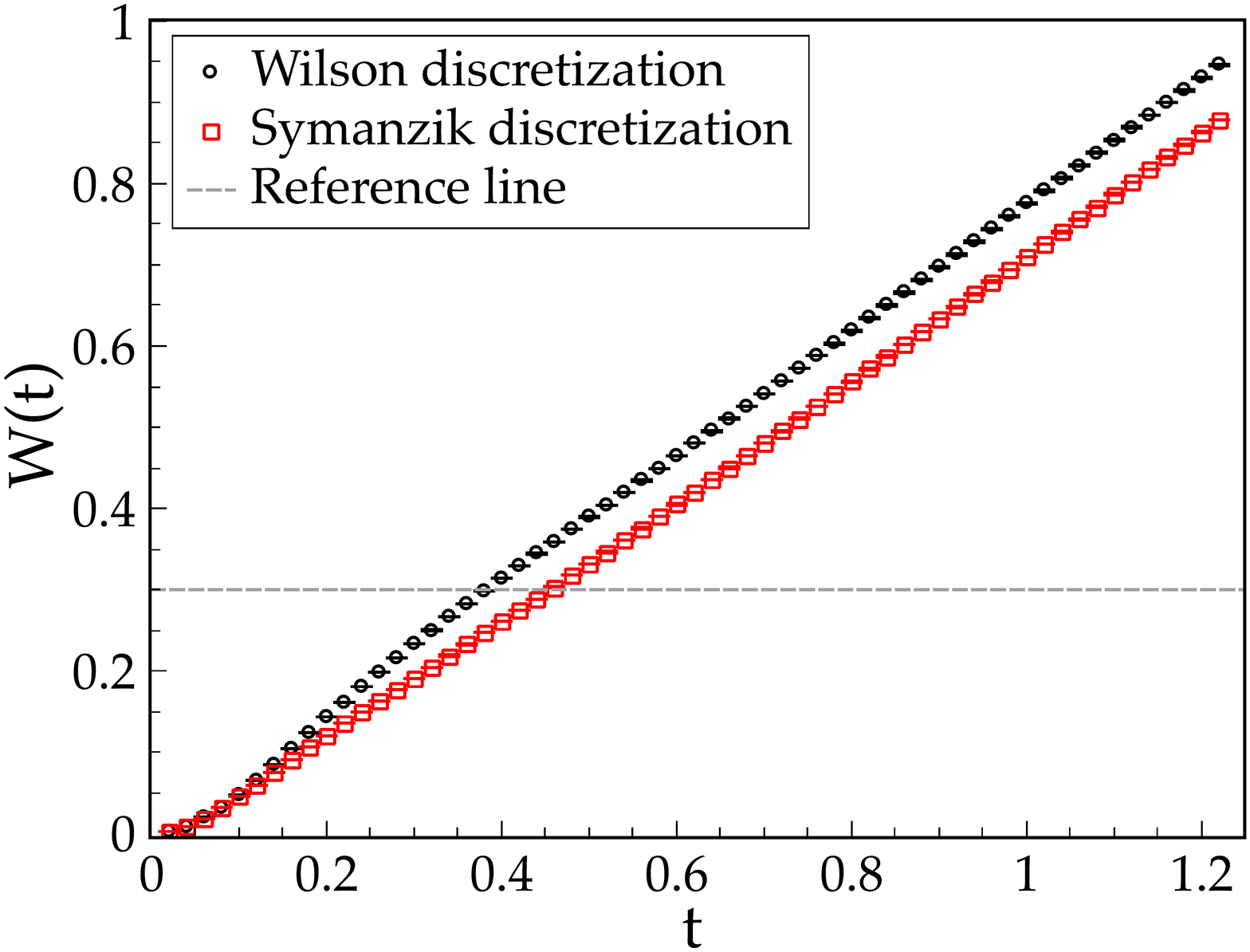}
\hfill
\includegraphics[width=.45\textwidth,origin=c]{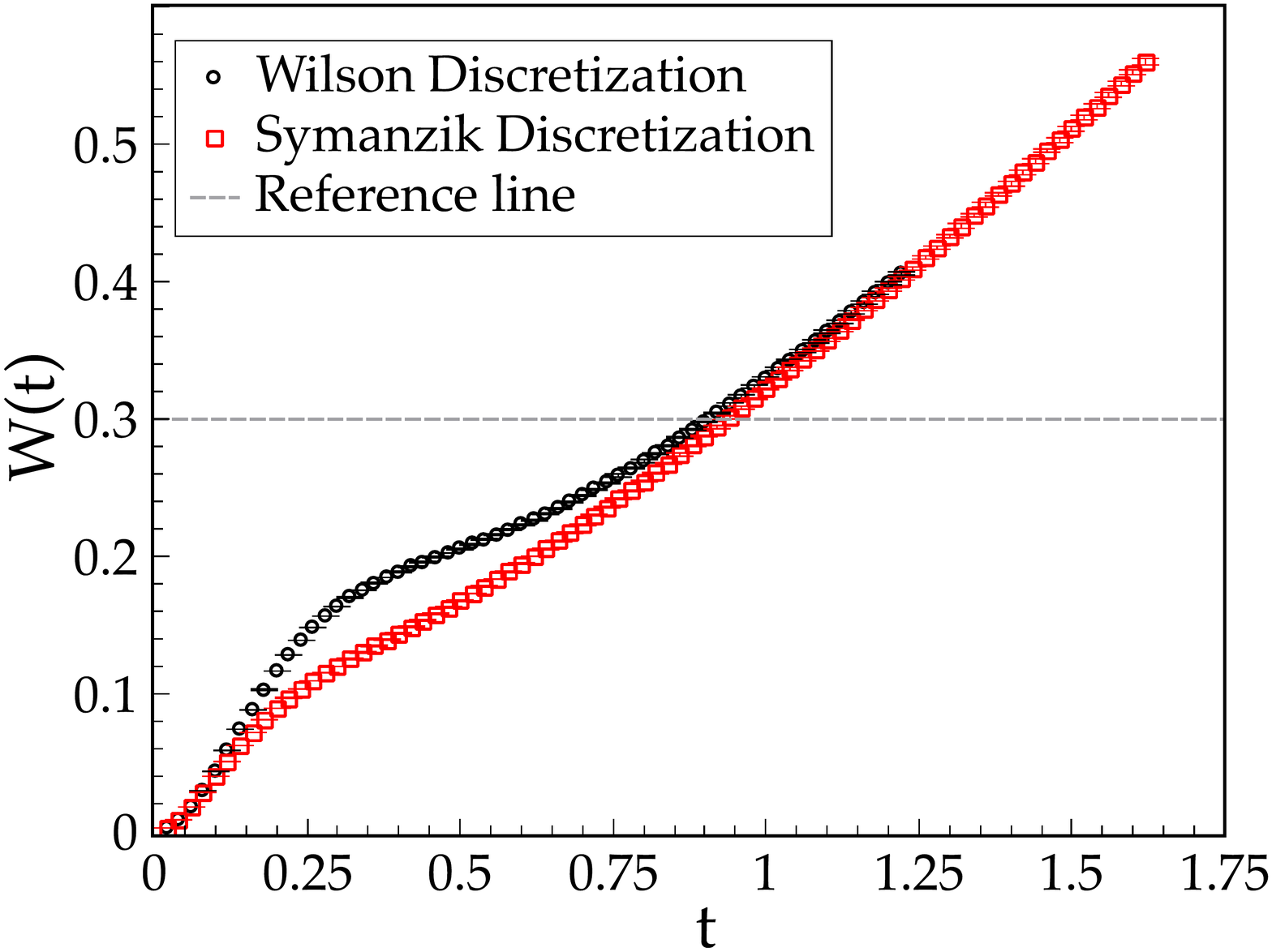}
\caption{ Plots of $W(t)$ for the ensembles $N_f = 8, \beta = 4.1125$ (left) and $N_f=8, \beta = 4.275$ (right). The flows were measured with both Wilson (black circles) and Symanzik (red squares) discretizations.}
\label{fig:flows-nf8}
\end{figure}

Next, we present our preliminary results on the gradient flow and the associated measurements of $w_0$. Our results for both the Wilson and Symanzik flows are presented on Figure~\ref{fig:flows-nf6} and Figure~\ref{fig:flows-nf8} for our ensembles with $N_f = 6$ and $N_f=8$, respectively. As can be seen from both figures, the use of the Symanzik discretization helps diminish the discretization effects at small flow times. Also as expected, the ensembles at the critical $\beta$s obtained at $N_t =8$ exhibit a much better agreement between the Wilson and Symanzik flows at the point where $w_0$ is measured, them being closer to the continuum. Note, however, that here we do not attempt to perform a continuum extrapolation of our results. We must measure the normalization quantity at the critical couplings $\beta$s obtained from our finite temperature simulations. In order to move closer to the continuum, one would need to obtain the critical $\beta$s from finite temperature simulations with a larger $N_t$.

Table~\ref{table:flowResults} summarizes our results for the four ensembles for $w_0$. The corresponding results for $T_cw_0$ are plotted in Figure~\ref{fig:tcRatios} (right). There, we also include preliminary measurements obtained from a $N_f = 0$ (quenched) ensemble. The results in this case exhibit a clearly decreasing trend, which is not present for the string tension. However, the slope is not large enough to capture the vanishing of the chiral phase boundary before the loss of asymptotic freedom. Two possible explanations are that the $w0$ determined in this work is not UV enough and/or finite mass effects.  

\begin{table}[htb]
\centering
\begin{tabular}{| c | c | c | c | c |}
	\hline
	 & $N_t = 6$ (Wilson) & $N_t=6$ (Symanzik) & $N_t = 8$ (Wilson) & $N_t =8$ (Symanzik)  \\ \hline
	 $N_f = 6$ & $ 0.7751(2) $ & $0.8289(3)$  & 1.1199(13)& 1.1187(11)\\
	 $N_f = 8$ & $ 0.6184(1) $ & $0.6753(2)$ & 0.9520(15)& 0.9680(12)\\
	 \hline
\end{tabular}
\caption{\label{table:flowResults} Value of $w_0$ obtained from the four ensembles analyzed, for both the Wilson and Symanzik discretizations. }
\end{table}

\section{Conclusions and Outlook}

We have presented our preliminary results on the study of the preconformal region of the phase diagram of the $SU(3)$ gauge theory for a varying number of fundamental flavors. We are studying the chiral phase transition temperature normalized by two different quantities: the string tension and $w_0$. Our new results for $T_c/\sqrt{\sigma}$ indicate a flattening of the ratio for increasing $N_f$. On the other hand, $T_cw_0$ presents a diminishing but mild trend for increasing $N_f$. These observations might be related to a genuine scale separation, but more work is needed, in particular to understand the effect of a finite mass. Work on this project is ongoing. Studying these different ratios might shed a light on the genuine singularities (if any) associated with the emergence of conformality and the different scales in the pre-conformal region of the phase diagram; they can also provide useful information for holographic models. 

\section{Acknowledgements}

We thank Marc Wagner for providing us the code for the measurements of the Wilson loops. This work was in part based on the MILC Collaboration's public lattice gauge theory code, see {\tt http://www.physics.indiana.edu/?sg/milc.html}. The configurations used in the study were generated at the BlueGene/Q in CINECA, Italy. The Wilson loop measurements were carried out on the high-performance computing system $\varphi$ at KMI, Nagoya. Measurements of the gradient flow were performed at the Millipede Cluster at University of Groningen. This work is part of the research programme of the Foundation for Fundamental Research on Matter (FOM), which is part of the Netherlands Organization for Scientific Research(NWO).


\begin{thebibliography}{99}

\bibitem{Caswell:1974gg}
  W.~E.~Caswell,
  \emph{Phys.\ Rev.\ Lett.}\  {\bf 33} (1974) 244.

\bibitem{Lombardo:2014mda}
  M.~P.~Lombardo, K.~Miura, T.~J.~N.~da Silva and E.~Pallante,
  \emph{Int.\ J.\ Mod.\ Phys.}\  {\bf 29} (2014) 1445007.

\bibitem{Miransky:1996pd}
  V.~A.~Miransky and K.~Yamawaki,
  \emph{Phys.\ Rev.\ D} {\bf 55} (1997) 5051
   [Erratum-ibid.\ D {\bf 56} (1997) 3768].

\bibitem{Appelquist:1998rb}
  T.~Appelquist, A.~Ratnaweera, J.~Terning and L.~C.~R.~Wijewardhana,
  \emph{Phys.\ Rev.\  D} {\bf 58}, 105017 (1998).

\bibitem{Braun:2010qs}
  J.~Braun, C.~S.~Fischer and H.~Gies,
  \emph{Phys.\ Rev.\ D} {\bf 84} (2011) 034045;
  %
  J.~Braun and H.~Gies,
  \emph{JHEP} {\bf 1005} (2010) 060;
  J.~Braun and H.~Gies,
  \emph{JHEP} {\bf 0606} (2006) 024.

\bibitem{Borsanyi:2012zs}
  S.~Borsanyi, S.~Durr, Z.~Fodor, C.~Hoelbling, S.~D.~Katz, S.~Krieg, T.~Kurth and L.~Lellouch {\it et al.},
  \emph{JHEP} {\bf 1209} (2012) 010.

\bibitem{Miura:2012zqa}
  %
  A.~Deuzeman, M.~P.~Lombardo and E.~Pallante,
  \emph{Phys.\ Lett.\ B} {\bf 670} (2008) 41;
  %
  K.~Miura, M.~P.~Lombardo and E.~Pallante,
  \emph{Phys.\ Lett.\ B} {\bf 710} (2012) 676;
  %
  K.~Miura and M.~P.~Lombardo,
  \emph{Nuclear Physics B} {\bf 871} (2013) 52.

\bibitem{Luscher:2010iy}
  M.~Luscher,
  \emph{JHEP} {\bf 1008} (2010) 071.

\bibitem{Luscher:2009eq}
  M.~Luscher,
  \emph{Commun.\ Math.\ Phys.}\  {\bf 293} (2010) 899.
  
\bibitem{Miura:2014aqa}
  K.~Miura, A.~Deuzeman, M.~P.~Lombardo, T.~N.~da Silva and E.~Pallante,
  \emph{PoS LATTICE} {\bf 2013} (2013) 074.
    
\bibitem{Laermann:1996xn}
  E.~Laermann,
  \emph{Nucl.\ Phys.\ A} {\bf 610} (1996) 1C.
  
\bibitem{Karsch:2000kv}
  F.~Karsch, E.~Laermann and A.~Peikert,
  \emph{Nucl.\ Phys.\ B} {\bf 605} (2001) 579.
  
\bibitem{Engels:1996ag}
  J.~Engels, R.~Joswig, F.~Karsch, E.~Laermann, M.~Lutgemeier and B.~Petersson,
  \emph{Phys.\ Lett.\ B} {\bf 396} (1997) 210.
  

\end{thebibliography}
\end{document}